# Open problems in ageing science: A roadmap for biogerontology

Angelo Talay [1], Aleksey V. Belikov [1], Paul Ka Po To [1], Hamid H. Alfatemi [1], Uri Alon [2], Joris Deelen [3,4,5], Collin Y. Ewald [6], David Gems [7], Vera Gorbunova [8], Jan Gruber [28,29], Sara Hägg [9], John Hemming [1], Steve Horvath [10], Alaattin Kaya [11], Caitlin J. Lewis [12], Andrea Maier [13,14], Maria B Marinova [24], Graham Pawelec [15,16], Shahaf Peleg [17], Suresh Rattan [18], Morten Scheibye-Knudsen [27], Tomas Schmauck-Medina [19], Vardan Saroyan [1], Andrei Seluanov [20], Alexandra Stolzing [21], Emma Teeling [22], Robert W. Williams [23], Todd White [24], Maximilian Unfried [24,25,26], João Pedro de Magalhães [1,*]

## Affiliations

[1] Genomics of Ageing and Rejuvenation Lab, Institute of Inflammation and Ageing, University of Birmingham, Birmingham, United Kingdom.
[2] Department of Molecular Cell Biology, Weizmann Institute of Science, Rehovot, Israel.
[3] Molecular Epidemiology, Department of Biomedical Data Sciences, Leiden University Medical Center, 2333 ZC, Leiden, Netherlands.
[4] Max Planck Institute for Biology of Ageing, 50931 Cologne, Germany.
[5] Cologne Excellence Cluster on Cellular Stress Responses in Ageing-Associated Diseases (CECAD), University of Cologne, 50931 Cologne, Germany.
[6] Laboratory of Extracellular Matrix Regeneration, Institute of Translational Medicine, Department of Health Sciences and Technology, ETH Zürich, Schwerzenbach CH-8603, Switzerland.
[7] University College London, Institute of Healthy Ageing, London, United Kingdom.
[8] Department of Biology, University of Rochester, Rochester, NY 14627, USA.
[9] Department of Medical Epidemiology and Biostatistics (MEB), Karolinska Institutet, Stockholm, Sweden.
[10] Altos Labs, Cambridge, United Kingdom.
[11] University of Virginia, School of Medicine. Dpt. Of Biochemistry and Mol. Genetics, Charlottesville, VA, USA 22908.
[12] Longevity Escape Velocity Foundation.
[13] Department of Human Movement Sciences, @AgeAmsterdam, Faculty of Behavioural and Movement Sciences, Amsterdam Movement Sciences, Vrije Universiteit, Van Der Boechorststraat 7, 1081 BT, Amsterdam, Netherlands.

[14] NUS Academy for Healthy Longevity, Yong Loo Lin School of Medicine, National University of Singapore, MD11 Clinical Research Centre, #03-01, 10 Medical Drive, Singapore 117597, Singapore.
[15] Institute of Immunology, University of Tuebingen, Tuebingen, Germany.
[16] Health Sciences North Research Institute, Sudbury, ON, Canada.
[17] Research Group Epigenetics, Metabolism and Longevity, Institute for Farm Animal Biology, 18196 Dummerstorf, Germany.
[18] Department of Molecular Biology and Genetics, Aarhus University, Aarhus, Denmark.
[19] Department of Clinical Molecular Biology, University of Oslo and Akershus University Hospital, Lørenskog, Norway.
[20] Department of Medicine, University of Rochester Medical Center, Rochester, NY 14627, USA.
[21] Centre for Biological Engineering, Wolfson School of Mechanical, Electrical and Manufacturing Engineering, Ashby Rd, Loughborough LE11 3TU, UK.
[22] School of Biology and Environmental Science, University College Dublin, Ireland.
[23] Department of Genetics, Genomics and Informatics, University of Tennessee Health Science Center; Memphis, TN, USA.
[24] The Thalion Initiative, 177 Huntington Avenue, 17th Floor Boston, MA 02115.
[25] Healthy Longevity Translational Research Program, Yong Loo Lin School of Medicine, National University of Singapore, Singapore, Singapore
[26] Department of Biochemistry, Yong Loo Lin School of Medicine, National University of Singapore, Singapore, Singapore
[27] Center for Healthy Aging, Department of Cellular and Molecular Medicine, University of Copenhagen, Denmark
[28] Healthy Longevity Translational Research Program, Yong Loo Lin School of Medicine, National University of Singapore
[29] Center for Healthy Longevity, National University Health System, Department of Biochemistry, Yong Loo Lin School of Medicine, National University of Singapore
*Corresponding author: jp@senescence.info

## Abstract

The field of ageing science has gone through remarkable progress in recent decades, yet many fundamental questions remain unanswered or unexplored. Here we present a curated list of 100 open problems in ageing and longevity science. These questions were collected through community engagement and further analysed using Natural Language Processing to assess their prevalence in the literature and to identify both well-established and emerging research gaps. The final list is categorized into different topics, including molecular and cellular mechanisms of ageing, comparative biology and the use of model organisms, biomarkers, and the development of therapeutic interventions. Both long-standing questions and more recent and specific questions are featured. Our comprehensive compilation is available to the biogerontology community on our website ([www.longevityknowledge.app)](www.longevityknowledge.app)). Overall, this work highlights current key research questions in ageing biology and offers a roadmap for fostering future progress in biogerontology.

## Introduction

Ageing and longevity science, dedicated to extending and understanding the lifespan and healthspan of organisms, has its roots in gerontology which emerged as a distinct discipline in the early 20th century [1]. Given the complexity of the ageing phenotype, there was initial scepticism about whether the ageing process could be effectively studied or intervened upon through scientific examination [2]. A significant breakthrough came in 1935 when McCay et al [3] demonstrated that dietary restriction could modulate lifespan in rats. This finding, later observed in multiple other organisms, suggested the existence of conserved pathways that could extend lifespan [4,5]. Since then, gene variants influencing lifespan have been identified in various organisms, including genes in metabolic pathways like the insulin/IGF-1 signalling pathway [6]. Current applications of these research advancements have led to several promising pharmacological interventions [5]. In addition to pharmacological approaches, recent advances in biomarkers of ageing have led to the development of epigenetic clocks [7]. While the underlying mechanisms of ageing remain debatable, several attempts have been made at developing frameworks that catalogue potential ageing processes [8,9].

Despite these advancements, the field of longevity science is at a crucial point as it continues to face numerous open problems that hinder further progress. Recent works have highlighted fundamental knowledge gaps [10] and strong disagreements amongst scientist studying ageing [11,12]. Addressing these challenges is critical for unlocking new insights and developing effective interventions to extend both lifespan and health span. In 1977, Strehler's seminal work, "Time, Cells, and Aging," presented a list of pivotal questions about ageing at that time, highlighting the complexity and the need for sustained research and collaboration [13]. Indeed, asking the right questions is essential for scientific research and, specifically, for steering the future directions of ageing and longevity science.

Building on the work of Strehler, we now present a new list of 100 open problems in ageing science, identified and curated through a combination of community engagement and text-mining approaches. These problems span a wide range of topics, from molecular biology and comparative approaches to translational efforts and clinical applications. By outlining these 100 problems, we aim to guide and provide goals for future research and map the key areas where knowledge gaps exist.

These open problems are presented on our website (https://longevityknowledge.app), where users can interact with and find more information on each selected problem.

3# Methods

### Compilation of open problems

First, we developed an initial website and database where scientists and users could submit their open problems. Input was garnered through advertising the website on social media and mailing lists to the research community; the social media platforms used were Twitter/X, Facebook and LinkedIn. Users were able to submit a title and description of their open problem, which was then saved to our database for downstream analysis.

In addition to creating a platform to gather open problems, we held a three-day workshop in Birmingham (UK) in collaboration with the Thalion Initiative, hosting 24 scientists in the ageing field. This workshop aimed to foster collaboration and develop current research questions in need of addressing. Throughout the workshop, attendees were grouped to generate ideas for open problems. After discussions, each group presented their ideas for feedback and to generate further ideas. All generated open problems were recorded into our database.

As we aimed to create a list of 100 open problems, we first narrowed down our obtained open problems, where we removed poorly formulated open problems and those deemed non-relevant to the field as well any duplicated open problems.

### Literature-Driven Analysis of Open Problems Using NLP

To help refine the final list of open problems, we employed Natural Language Processing (NLP) techniques to analyse the representation of each open problem in the existing literature on ageing. This analysis helped identify the relevance and research focus of each open problem within the field. By linking the open problems to indexed articles in PubMed, we aimed to gather insights into which questions were already well-represented in the literature, and which might warrant further investigation.

First, we collected all the abstracts and articles in the PubMed database under the MeSH Term "Ageing" in the years ranging from 1963 to 2023 and combined them. Articles without available abstracts were excluded from the comparison. Then we generated numerical representations (embeddings) for all the open problem titles and for all the articles with their combined title and abstract. This was done using PubMedBert [14], a language model trained on text from the PubMed database, developed for various NLP tasks. Using these embeddings we calculated cosine similarity scores, ranging from -1 to 1, to quantify how closely an open problem title aligned with an article. To filter out unrelated pairings, we removed any open problem-article pairs that scored 0.2 and below.

33

Following this, we then applied a cross-encoder model, Med-CPT, another language model trained on PubMed search logs, for the task of query-article matching [15]. We applied this model on the remaining pairs of open problems and articles. The output returned by the model was logits, which we then converted into probabilities between 0 and 1 using a sigmoid function. To help us choose the relevant pairings, we also applied a natural learning inference (NLI) model based on PubMedBert. This model assessed the relationship between each open problem and article and applies a label that determines whether the article supports, contradicts or is unrelated to the open problem title. Finally, pairs with both a probability score of at least 0.8 and a supporting NLI label were kept. The number of articles for each open problem title was then counted, which provided us a measure of how frequently the topic appeared in scientific literature.

### Grouping and Thematic Analysis of Open Problems

To group the open problems, we used the previously calculated embeddings of the open problem titles and then calculated a cosine similarity score between only the titles and stored them as a matrix. Using these scores we employed consensus clustering, a method that combined multiple clustering techniques to select the most consistent group structure across all clustering methods. The clustering methods used were K-means clustering, agglomerative clustering and Gaussian Mixture Models. The groupings provided by the consensus clustering gave us a starting point for grouping them into themes. However, given the complexity and subjectivity of the topics, we also relied on manual curation to ensure each grouping was appropriate.

The resulting dataset comprised all the open problems along with their associated article counts, organised into initial groups based on similar themes. Where appropriate, we merged highly similar questions into a single entry. From this dataset, João Pedro de Magalhães selected the final list of 100 open problems, prioritising those deemed important while ensuring a diversity of topics, using the initial groupings and article counts as a complementary guide. In other words, we aimed to have a broad range of questions in terms of topics and article counts.

The final 100 open problems are published on our website: www.longevityknowledge.app.

### Results

To help identify and prioritise the more pressing open problems in longevity and ageing science we combined community input, data driven analysis and manual review (see Methods). All open problems were collected through an online platform as well as a dedicated workshop which ensured engagement from the scientific community. We assessed the open problems by applying NLP techniques to analyse their prevalence in the PubMed database. This allowed us to quantify the extent to which an open problem may have been explored, highlight potential gaps in research and inform us on the final selection of open problems. Alongside this we utilised clustering methods to help us group our open problems into appropriate categories for easier exploration for users on our website.

## Data collection of open problems

The number of open problems we collected through submissions via the website and through the workshop totalled to 290. Online submissions totalled to 160 open problems while the workshop produced 130. Initial pre-filtering of questions before grouping and NLP analysis reduced this to a total of 204. These 204 open problems were then used for clustering and NLP analysis against PubMed articles.

## NLP Analysis of Open Problems Using PubMed Literature

Collection of all PubMed articles under the MeSH term of "Ageing" totalled to 389,627 articles. After the removal of articles that did not contain an abstract, the number of unique articles totalled to 200,228 articles which were then paired to the 204 selected open problems. This created a total of 40,846,512 pairs of open problems and articles to be analysed. Using the PubMedBert embeddings of the open problems and article texts we assessed the relationship between the open problems and the PubMed ageing articles.

The NLP analysis of the 204 open problems revealed variability in their representation within the ageing research literature (Supplementary Table 1). After the application of the language models, a total of 172,031 relevant article and open problem pairs were identified. The representation of individual open problems ranged from 1 to 10,808 articles, with a median of 437 and a mean of 847.4 articles per problem. The standard deviation of 1,176.7 reflects substantial disparities in research focus across different open problems in ageing research.

**Summary of the Top and Bottom Open Problems**

A statistical analysis of the top and bottom 20 open problems ranked by the total number of associated articles shows a disparity in their representation within the ageing research literature (Supplementary Table 1). The top 20 open problems collectively account for 69,322 articles, representing 40.3% of the total dataset, with an average of 3,466.1 articles per problem. In contrast, the bottom 20 problems account for only 341 articles, 0.2% of the dataset, with an average of just 17.05 articles per problem.

The top 20 problems dominate the literature, reflecting well-established questions regarding the causes of ageing and even whether fundamental ageing processes exist (Table 1). These problems align closely with current research priorities and methodologies, such as cellular and molecular mechanisms of ageing, evolutionary and comparative biology, the role of model organisms in ageing research, biomarkers, and developing interventions, reinforcing their centrality to the field. Article counts for these problems range from 2,196 to 10,808, showing the breadth of interest in these topics.

In contrast, the bottom 20 open problems have article counts ranging from 1 to 36. These less explored problems often focus on emerging or niche areas, including methodological challenges, novel therapeutic approaches, and unexplored biological mechanisms (Table 2). Naturally, more underrepresented open problems tend to be less supported by existing literature. Nonetheless, these often more specific topics can provide tangible research goals in a shorter time frame (Table 2).

| **Category** | **Open Problem** | **Articles (n)** |
|---|---|---|
| **Broad Fundamental Questions** | Why do we age? | 10,808 |
| | Do fundamental ageing processes exist and if so, how do they synchronize ageing changes and promote age-related diseases? | 2,366 |
| | Which tissues, organs or cell types contribute more to ageing? | 2,304 |
| **Cellular and Molecular Mechanisms** | Does somatic mutation accumulation cause ageing? | 5,977 |
| | What molecular and cellular processes modulate the pace of ageing in mammals? | 3,885 |
| | Which and when are senescent cells beneficial and detrimental? | 2,280 |
| **Evolutionary and Comparative Biology** | What mechanisms determine the longevity of long-lived species? | 3,668 |
| **Model Organisms** | Which ageing changes in model organisms also change in a similar way in humans? | 3,110 |
| **Biomarkers** | How can we measure intrinsic biological age in individuals and translate this knowledge into accurate biomarkers of ageing? | 2,545 |
| **Interventions** | How can we reverse or restore cell function lost during ageing? | 2,196 |

**Table 1:** Examples of open problems with significant representation in the literature, categorised by topic.

| Category | Open Problem | Articles (n) |
|---|---|---|
| **Biomarkers** | How can we measure the extent and pace of changes in the homeodynamic space during ageing? | 1 |
| **Novel Therapeutic Approaches** | How can we break through the current human lifespan ceiling of 122 years? | 4 |
| | How and which interventions should be prioritized for human clinical trials? | 8 |
| | Can a combination of senolytics with ROCK inhibitors and 5-LOX inhibitors contribute to tissue rejuvenation? | 12 |
| | Can we apply stem cell therapy based on our own young stem cells (e.g., from cord blood) to modulate ageing? | 19 |
| | Could blood clean-up be used to target ageing processes? | 32 |
| **Specific Mechanisms** | How much does the immune response to mutated cells or altered (oxidized, misfolded) molecules (e.g. oxLDL, beta amyloid, alpha synuclein) contribute to ageing? | 9 |
| | How much do positive feedback loops and chain reactions contribute to ageing? | 23 |
| | How many diseases of ageing are caused by the trapping of citrate in the mitochondria? | 36 |

**Table 2**: Examples of open problems with the least representation in literature, categorised by topic.

## Final Themes and Distribution of Open Problems

The final list of 100 open problems was grouped into 11 themes, ensuring thematic alignment. This grouping process facilitated the organisation of problems into a logical structure that reflects the breadth of the open problems. Each problem was assigned to one primary theme that best fitted its focus.

Figure 1 presents the distribution of the 100 open problems across the 11 themes. The largest proportions were assigned to broader Ageing Mechanisms, more specific Molecular Mechanisms, and Interventions, which collectively accounted for over half of the selected problems. Themes such as Environmental and Physical Factors and Diversity in Human Ageing were less represented, which may reflect they are less explored topics in the field. The final list of 100 open problems is available as Supplementary Table 2.

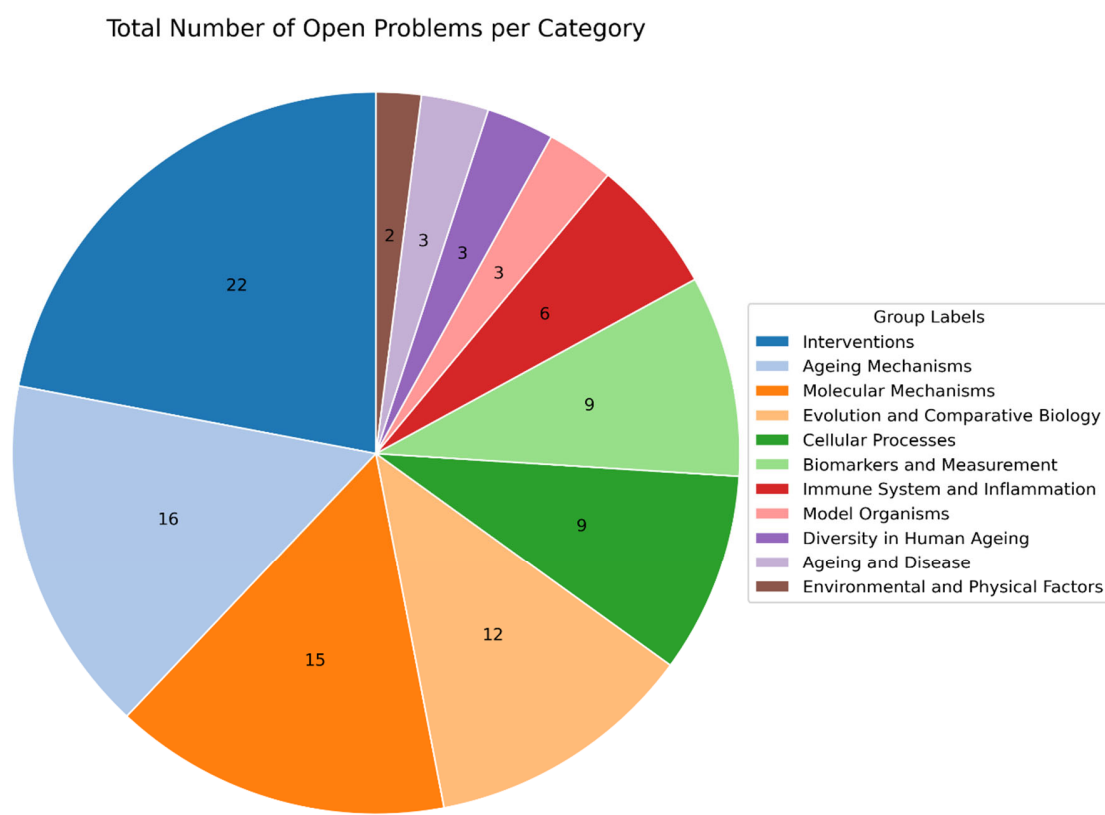

**Figure 1:** Pie chart of the total number of open problems distributed across the 11 themes.

**Presentation of the Final 100 Open Problems**

To maximise engagement and accessibility, the final list has been published on the **Longevity Knowledge App website** (www.longevityknowledge.app). The website serves as an interactive platform, enabling researchers and the public to explore the open problems and their themes. Each problem is presented with detailed information, including:

- Title of the open problem.
- Theme to which it belongs.
- Metadata, such as genes or compounds, that link to external databases.
- Links to related open problems.
- A section where users can post any solutions or proposals that can tackle the open problem.

Our website is designed to stimulate further discussion, collaboration, and roadmap planning in the ageing research scientific community.

# Discussion

Asking the right questions is crucial to advance a scientific field and steer its future directions. As such, we took inspiration from the pioneering work of Strehler to identify, through a collective and systematic effort, a new set of 100 open problems in ageing science. Our list of open problems purposely includes a combination of broad, big-picture questions and specific questions. While broad questions, for example regarding why we age, set long-term prospects in the field, more specific questions may be answerable in the foreseeable future. In other words, we need both big picture questions – for direction – and precise questions to advance biogerontology, which our work provides. Besides, while some topics are more represented than others, for example questions on mechanisms of ageing are the most popular, we also aimed to have a diverse set of topics represented in our final list of questions. As such, our open problems will stimulate further discussions and provide a cornerstone in ageing research for years to come.

The topics covered in our open questions reflect those we received on our website and the discussions during our workshop. Not surprisingly, a major emphasis is on mechanisms of ageing and understanding why we age. This is still a key but major open question in the field as the drivers of ageing remains open to debate [16]. Many open questions focus on processes or mechanisms hypothesized to be associated with ageing, again ranging from broad questions to more specific ones that are amenable to experimentation. We hope some of our more specific questions broaden the field's research directions and are useful for students and researchers to address in the next five years.

Another broad topic of great interest is developing interventions targeting ageing. Again, there is significant discussion concerning longevity therapeutics, regarding both the effectiveness of existing potential therapies and how to test them in a clinical setting. It is clear, however, that there is great interest in developing effective interventions for ageing and this is one of the major areas for demonstrating clinical efficacy in the future [5]. Several questions also concern biomarkers, e.g. which would be the most suitable for evaluating therapies. Questions also arose regarding the nature of existing biomarkers, such as epigenetic clocks, the genetic and environmental determinants of ageing in humans and model organisms and questions regarding recent methods such as partial reprogramming. A few questions concerned the nature of the ageing process, which others have also debated without reaching a consensus [11,12], for example, regarding cell autonomous and systemic contributors to ageing across tissues and organs. A subset of questions also concerned specific organs and tissues, such as the immune system, and the relationship between ageing processes and disease.

Evolution and comparative biology were another broad topic of interest, again a historically important topic that has not been resolved. Broadly speaking, we still do not understand species differences in ageing, and therefore it remains another major open area for research. Despite recent progress in finding common ageing markers across mammalian species [17], the level of conservation of ageing mechanisms across species remains under debate.

We did not use the initial set of open questions by Strehler, now nearly 50 years old [13], as a basis for our work. Hence, how do our 100 open questions compare to those of Strehler? Broadly speaking, some topics remain the same, such as the role of genetics in ageing and longevity, how transcription and translation relate to ageing, the timing of development in long-lived species, species differences in ageing, and even gene therapies for ageing. A few of the questions from Strehler have been partially addressed, for example regarding the genetics of longevity in animal models [6], but most remain open. There is also an overlap in mechanisms of ageing between Strehler's list and ours, for example regarding mitochondria, cellular changes, and the role of the immune system. In terms of differences, Strehler gives much more emphasis to CNS diseases, including not only neurodegenerative conditions but also others such as dyslexia and Batten's disease, which are absent from our list. Furthermore, our list is shaped by modern priorities, including a strong emphasis on developing and evaluating interventions—such as senolytics, partial reprogramming, and establishing biomarkers to guide clinical application. This shift may reflect a broader evolution in the field, from characterizing ageing changes to actively seeking to modulate them. Our list also reflects a deeper knowledge of molecular biology. For example, Strehler referred to enzymes, tRNAs and ribosomes—broader aspects of biology that are rarely now studied in the context of ageing, even though their role remains an open question. With the advent of molecular biology and large-scale -omics analyses, we should perhaps take notice of old-fashioned questions for which we still lack answers. As such, we see our list as reflecting advances in the field, but complementary to those of Strehler. We have no doubt that future lists of open questions in biogerontology will only partly overlap with ours. While asking the right questions is essential to advance science, answering – or attempting to answer – them often leads to more questions.

In conclusion, our work aims to provide future directions for ageing research that will allow us to address key challenges that remain unanswered. Our open questions reflect current large discussions and unknowns in the field of biogerontology. In spite of progress, for example in developing interventions that retard ageing in preclinical models, there is still huge debate regarding the underpinning mechanisms of ageing, the nature of biomarkers of ageing and a route to developing effective interventions for ageing in humans. Our questions reflect current key disagreements in the field (Table 1), and we hope will serve as inspiration and guide to researchers as well as lay a path for advancing biogerontology.

## Acknowledgements

We thank all the workshop participants and website contributors. Further thanks to Lada Nuzhna for assistance and advice throughout the project. We are also grateful to current and past members of the Genomics of Ageing and Rejuvenation Lab for valuable discussions. The project was supported by a Longevity Impetus Grant and the Hevolution Foundation. We are also grateful to the Thalion Initiative for supporting the workshop in Birmingham. The Genomics of Ageing and Rejuvenation Lab is further supported by LongeCity and the Biological Sciences Research Council (BB/V010123/1).

## Conflict of interest

JPM is CSO of YouthBio Therapeutics, an advisor/consultant for the BOLD Longevity Growth Fund and NOVOS, and the founder of Magellan Science Ltd, a company providing consulting services in longevity science. SH is a founder and paid consultant of the non-profit Epigenetic Clock Development Foundation and a Principal Investigator at the Altos Labs, Cambridge Institute of Science, a biomedical company that works on rejuvenation.

## Data Availability Statement

The code used to support the findings of this study is available at:
https://github.com/angelotalay/opl-analysis

12## Bibliography

1.  National Research Council (US), C.o.C.T.a.A. *Aging in Today's Environment*, 234 (The National Academies Press, Washington, DC, 1987).

2.  Partridge, L. The new biology of ageing. *Philosophical Transactions of the Royal Society B: Biological Sciences* **365**, 147-154 (2010).

3.  McCay, C.M., Crowell, M.F. & Maynard, L.A. The Effect of Retarded Growth Upon the Length of Life Span and Upon the Ultimate Body Size: One Figure. *The Journal of Nutrition* **10**, 63-79 (1935).

4.  Fontana, L., Partridge, L. & Longo, V.D. Extending healthy life span--from yeast to humans. *Science* **328**, 321-6 (2010).

5.  de Magalhaes, J.P., Wuttke, D., Wood, S.H., Plank, M. & Vora, C. Genome-environment interactions that modulate aging: powerful targets for drug discovery. *Pharmacol Rev* **64**, 88-101 (2012).

6.  Kenyon, C.J. The genetics of ageing. *Nature* **464**, 504-12 (2010).

7.  Horvath, S. & Raj, K. DNA methylation-based biomarkers and the epigenetic clock theory of ageing. *Nat Rev Genet* **19**, 371-384 (2018).

8.  Kennedy, B.K. *et al.* Geroscience: linking aging to chronic disease. *Cell* **159**, 709-13 (2014).

9.  Lopez-Otin, C., Blasco, M.A., Partridge, L., Serrano, M. & Kroemer, G. Hallmarks of aging: An expanding universe. *Cell* **186**, 243-278 (2023).

10. Rattan, S.I.S. Seven knowledge gaps in modern biogerontology. *Biogerontology* **25**, 1-8 (2024).

11. Cohen, A.A. *et al.* Lack of consensus on an aging biology paradigm? A global survey reveals an agreement to disagree, and the need for an interdisciplinary framework. *Mech Ageing Dev* **191**, 111316 (2020).

12. Gladyshev, V.N. *et al.* Disagreement on foundational principles of biological aging. *PNAS Nexus* **3**, pgae499 (2024).

13. Strehler, B.L. XI - Some Unexplored Avenues of Cellular Aging—Current and Future Research. in *Times, Cells, and Aging (Second Edition)* (ed. Strehler, B.L.) 372-391 (Academic Press, 1977).
12

14. Gu, Y. *et al.* Domain-Specific Language Model Pretraining for Biomedical Natural Language Processing. *ACM Trans. Comput. Healthcare* **3**, Article 2 (2021).
15. Jin, Q. *et al.* MedCPT: Contrastive Pre-trained Transformers with large-scale PubMed search logs for zero-shot biomedical information retrieval. *Bioinformatics* **39**(2023).
16. de Magalhães, J.P. Distinguishing between driver and passenger mechanisms of aging. *Nat Genet* **56**, 204-211 (2024).
17. Lu, A.T. *et al.* Universal DNA methylation age across mammalian tissues. *Nat Aging* **3**, 1144-1166 (2023).